\documentstyle[floats,pre,aps,galley,psfig]{revtex}	% galley

  \newlength{\figwidth}
  \newlength{\refspace}

\begin{document}
\draft

%%%%%%%%%%%%%%%%%%%%%%%%%%%%%%%%%%%%%%%%%%%%%%%%%%%%%%%%%%%%%%%%%%%%%%%%%%
%% Title page
%%%%%%%%%%%%%%%%%%%%%%%%%%%%%%%%%%%%%%%%%%%%%%%%%%%%%%%%%%%%%%%%%%%%%%%%%%

%\Draft			% For "Draft" diagonal annotation in galley mode
\Preprint		% For "Preprint" diagonal annotation in galley mode

\title{Deformation of Small Compressed Droplets}
\author{Martin-D. Lacasse$^1$,
Gary S. Grest$^1$, and Dov Levine$^2$}
\address{%
$^1$Corporate Research Science Laboratories\\
Exxon Research and Engineering Co., Annandale, NJ 08801\\
$^2$Department of Physics, Technion, Haifa, 32000 Israel}
\date{March 11, 1996}

\maketitle
{
\begin{abstract}
We investigate the elastic properties of small droplets
under compression.
The compression of a bubble by two parallel plates is solved
exactly and it is shown that a lowest-order expansion of
the solution reduces
to a form similar to that obtained by Morse and Witten.
Other systems are studied numerically
and results for configurations involving between
2 and 20 compressing planes are presented.
It is found that the
response to compression depends on the number of planes.
The shear modulus is also calculated for common
lattices and the stability crossover between
f.c.c.\ and b.c.c.\ is discussed.
\end{abstract}
}
{
\pacs{PACS numbers: 82.70.Kj, 81.40.Jj, 62.20.Dc}
}

\narrowtext

%%%%%%%%%%%%%%%%%%%%%%%%%%%%%%%%%%%%%%%%%%%%%%%%%%%%%%%%%%%%%%%%%%%%%%%%%%
%% Core Text
%%%%%%%%%%%%%%%%%%%%%%%%%%%%%%%%%%%%%%%%%%%%%%%%%%%%%%%%%%%%%%%%%%%%%%%%%%

%%%% Example on how to include a figure in text.
%% \begin{figure}[tbp]
%% \centerline{\psfig{figure=fig1.ps,width=\figwidth}}
%% \caption{This is the caption. Bla
%% bla bla.
%% \label{fig1}}
%% \end{figure}

\section{Introduction}
\label{s:intro}

Emulsions consist of a mixture of two immiscible fluids,
one of which, generally an oil, is dispersed as small droplets
in the continuous phase of the other fluid, generally water.
The interfaces are
stabilized by a surfactant, preventing coalescence.
Emulsions are materials with quite unusual properties:
despite being comprised solely of fluids, they become elastic
solids when the droplets
are compressed to a large enough volume fraction $\varphi$
by extracting the continuous phase
by the application of an osmotic pressure $\Pi$.
The origin of the elasticity is the interfacial energy of the
droplets. At low volume fractions, the surface tension $\sigma$ ensures
that the droplets are spherical in shape. However, at higher $\varphi$, the
packing constraints force the droplets to deform, thus storing energy.
Consequently, it is a fundamental issue to determine the increase of the
surface area of a droplet resulting from an arbitrary deformation.

It is generally believed that the application
of a shear strain to a compressed
emulsion causes the droplets to further deform,
thereby storing
more energy~\cite{princen0,kraynik,bolton,morse,buzza}.
According to this picture,
the scale of both
the osmotic pressure,
and the elastic shear modulus $G$
is set by the Laplace pressure of the
droplets, $(2\sigma/R)$, where $R$ is the radius
of an undeformed droplet.
Recent experiments~\cite{mason}
on the elastic properties of compressed
monodisperse emulsions of silicone-oil in water
confirm the role of the Laplace pressure: the
experimental values of $G$
and $\Pi$ for different droplet
sizes, when scaled by $(\sigma/R)$,
collapse on a single curve~\cite{mason}.

Compression modifies the elastic properties
of a disordered emulsion, changing the response from
liquid-like to solid-like. The
onset of solid-like behavior is gradual; both $\Pi$ and $G$
increase smoothly from zero as the system is compressed
above $\varphi_c \approx 0.64$~\cite{mason},
the volume fraction at which disordered monodisperse
droplets are first deformed~\cite{berryman}.
For polydisperse emulsions, $\varphi_c$
is larger~\cite{princenka}, reflecting more efficient packing.
Two separate elastic responses of the
droplets are therefore of interest: to
direct compression and to shear.

The increase of the surface area of a droplet under compression comes from
the relatively low cost of deformation compared to the compressibility
of the internal fluid. Thus,
it is very reasonable to assume that the droplets
are composed of an incompressible fluid, and may be consequently
considered to have a fixed volume.
The effect of gravity may be nullified by density matching of the oil
and water, but in any case would be tiny because of the large surface
tension and small droplet size (typically of a $\mu$m)
of the experimental systems.

When the droplets are compressed just beyond the point they start
to deform, i.e.\ $\varphi \gtrsim \varphi_c$,
they form small facets separated by thin
surfactant-water-surfactant films.
Each of the facets created on the droplet surface has the
effect of {\it locally\/} decreasing the droplet
surface, but volume conservation
makes the droplet {\it globally\/} increase its
surface, and thus its energy. Due to
such non-local effects, the determination of the droplet response
to compression requires that it be considered as a whole. Consequently,
a complete analytic solution is limited to a few simple cases,
one of which is a drop squeezed between two parallel plates,
to be presented in Section~\ref{s:compression}.
The response to shear is still more difficult, but
fortunately, numerical tools capable of handling
such problems have recently become available~\cite{brakke}.

Upon more compression, droplets gradually take the shape
of a rounded polyhedron.
At this point, the water is contained within the connected
network of voids left between the oil droplets.
This network is made of thin veins, along which the facets meet,
and which connect larger regions located at the corners
of the polyhedra.
As $\varphi \rightarrow 1$, i.e.\ in the so-called dry foam limit,
mechanical equilibrium imposes constraints on these regions
which were first discussed by Plateau~\cite{plateau}.
In particular, exactly three films
meet at equal angles (120$^\circ$) along the veins
which become the edges of the polyhedra. Additionally,
the corner regions reduce to a point at which four edges
meet at equal tetrahedral angles.
These rules do not apply until most of the continuous phase
has been extracted.

The elasticity of emulsions has been the object
of numerous investigations over the last decade.
The first theoretical attempts~\cite{princen0,kraynik} were
for ordered lattices in two dimensions, which we shall briefly treat
in Section~\ref{s:2d}, mainly in order to contrast them
with three-dimensional systems. It should be noted that two-dimensional
systems are important in their own right, both
theoretically and experimentally. For example,
the effects of polydispersity and disorder
are largely studied in two dimensions~\cite{bolton,hutzler,durian}.

For real (three-dimensional) bubbles,
the only analytical expression~\cite{morse}
for the behavior of a droplet surface under compression
involves the application of an infinitesimal force at a single point.
A simplified approach consists in
assuming that each droplet becomes a truncated sphere
under compression~\cite{princen1}.
Using each of these approaches, the shear modulus of a simple
cubic (s.c.) lattice has been derived~\cite{buzza}.
These studies predict that the onset of the
shear modulus at $\varphi_c$ is discontinuous,
in contrast with the smooth quasi-linear increase of
$G$ at $\varphi_c$ found experimentally for disordered emulsions.

The relative importance of the disorder of the emulsions,
its polydispersity, and the response of the droplets themselves
is the object of current research~\cite{lacasse}.
This paper is concerned with the last:
we present the results of an investigation
of the response of an individual droplet to deformation.
In Section~\ref{s:2d} we present results for the 2-dimensional
case. Section~\ref{s:compression} derives exact results for
the compression of a droplet by two parallel plates. In the limit
of infinitesimal compression, our results are compared with
Morse and Witten's expression~\cite{morse}. For
a large range of compressions, we show that the energy has an anharmonic
power-law response to compression.
In Section~\ref{s:evolver}, we present numerical results for
the compression of a droplet in several local environments
using Brakke's Surface Evolver (SE) software~\cite{brakke}. We
again demonstrate that the energy of the squeezed droplet is
anharmonic, {\it with the power depending on the number of facets.}
Section~\ref{s:foam} describes the results obtained for a selection
of space-filling structures at the dry foam limit.
Section~\ref{s:shear} contains numerical calculations
of the droplet response to a shear deformation.
We discuss our results and conclude in Section~\ref{s:discussion}.

\section{response in two dimensions}
\label{s:2d}

Two-dimensional systems have characteristics not
found in three-dimensions. For example, monodisperse
arrays of disks order easily~\cite{bideau} in an hexagonal array
of surface fraction $\varphi_c = \pi/(2\sqrt{3}) \approx 0.9069$.
The elastic properties of two-dimensional ordered systems
have been discussed extensively by Princen~\cite{princen0,princen00}.
The response to deformation is also simpler in two dimensions.
A minimum free surface (a free curve in the present case)
is characterized by a uniform pressure
or, equivalently by a constant mean curvature. In two dimensions, the
surface is parameterized by only one radius of curvature and the minimum
free surface is therefore always an arc of a circle.

Let us consider the deformation of a two-dimensional droplet
of constant area confined inside a regular polygon of $n$ sides.
Only for $n = $ 3 ,4, and 6 can such polygons be tessellated,
but we shall consider general $n$, in particular to
contrast the two- and three-dimensional results. The ratio
of droplet area to polygon area is denoted by $\varphi$. Only
for the tessellating cases will $\varphi$
represent the surface covering fraction
of the corresponding lattice.
In the following, we assume that the droplet is non-wetting
so that the contact angles are zero.
At $\varphi_c = \frac{\pi}{n} \cot\frac{\pi}{n}$, the circular
droplet is undeformed and touches the polygon at
exactly $n$ points, being the midpoints of the polygon's edges.
If the polygon is now shrunk uniformly, the droplet will distort, and its
new minimum-perimeter shape will consist of small facets joined by circular arcs.
As $\varphi$ increases
above $\varphi_c$, the length of the flat portions increases and the
radius of curvature of the arcs decreases, reflecting increased droplet
pressure. At $\varphi = 1$, the radius of curvature of the arcs
becomes zero (infinite pressure), and the bubble takes on the
shape of the polygonal cell.

  \begin{figure}[tbp]
  \centerline{\psfig{figure=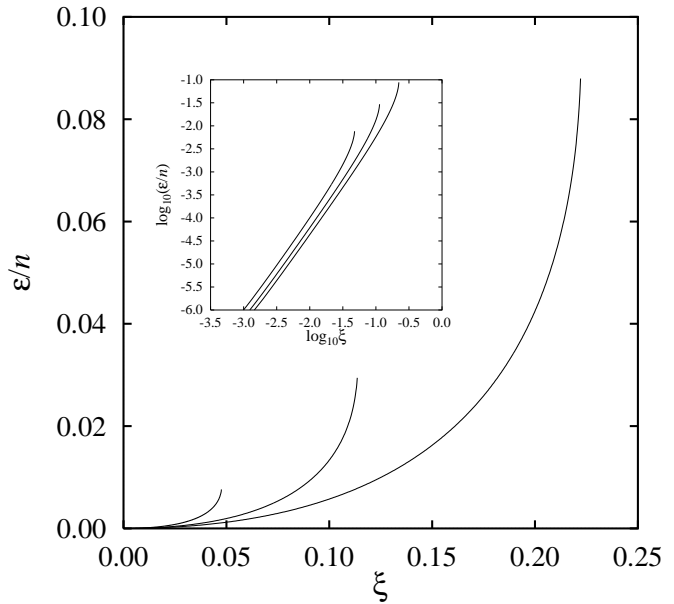,width=\figwidth}}
  \caption{The excess energy per facet of a two-dimensional droplet
  as a function of compression. The curves represent, from right to left,
  the tessellating cases $n$ = 3, 4, and 6. The insert shows the same
  data on a logarithmic scale.
  \label{af:2d}}
  \end{figure}

The interfacial energy of the droplets is the product of their line tension
$\sigma$ and the perimeter length $\ell$.
It is convenient to measure the excess energy by the following
dimensionless quantity
\begin{equation}
\varepsilon \equiv \frac{\ell}{2\pi R} - 1,
\end{equation}
where $R$ is the radius of the undeformed circular droplet.
Similarly, the degree of compression will be measured by the following
dimensionless displacement
\begin{equation}
\xi \equiv \frac{R - h}{R},
\label{ce:xidef}
\end{equation}
where $h$ is the perpendicular distance from the facets
to the center of the droplet.
$\xi$ is simply related to the volume
fraction through $\xi = 1 - \sqrt{\varphi_c/\varphi}$.
For any regular polygonal cell, $\varepsilon$ can be shown to be,
for $\varphi \geq \varphi_c$,
\begin{equation}
\varepsilon = \frac{1}{\sqrt{\varphi\varphi_c}}
- \sqrt{\frac{(1 - \varphi)(1 - \varphi_c)}{\varphi\varphi_c}} - 1.
\label{ae:energy}
\end{equation}
For $\varphi$ slightly greater than $\varphi_c$, we may expand the previous
expression in terms of $(\varphi-\varphi_c)$, and get
\begin{eqnarray}
\varepsilon & \approx & \left(\frac{1}{8 \varphi_c^2 (1-\varphi_c)}\right)
(\varphi-\varphi_c)^2,\nonumber\\
& \approx & \left(\frac{1}{2(1-\varphi_c)}\right) \xi^2.
\end{eqnarray}
Therefore, for small compression, the energy is harmonic, and
the prefactor depends on
the number of faces of the polygon through $\varphi_c$.
This is shown in Fig.~\ref{af:2d} where
the excess energy per facet is plotted for the tessellating cases
$n$ = 3, 4 and 6.  The end of each curve corresponds to
$\varphi \rightarrow 1$ for each lattice. The logarithmic
insert shows the wide range
over which harmonicity holds as well as the $n$ dependence of the prefactor
expressed by the different intercepts.

In two dimensions, the scaled osmotic pressure can be obtained from
\begin{equation}
\Pi/(\sigma/R) = 2 \varphi^2\frac{\partial \varepsilon}{\partial \varphi},
\end{equation}
and we find, using Eq.~\ref{ae:energy},
\begin{equation}
\Pi/(\sigma/R) = 
\left(\frac{\varphi}{\varphi_c}\right)^\frac{1}{2}
\left[\left(
\frac{1 - \varphi_c}{1 - \varphi}\right)^\frac{1}{2} - 1\right].
\label{ae:osmotic}
\end{equation}
This last expression, which depends on $n$ through
$\varphi_c$, is valid for all tessellations in two dimensions.
For $\varphi \gtrsim \varphi_c$, 
$\Pi$ can be shown to increase linearly with $(\varphi - \varphi_c)$.

In the model proposed by
Princen~\cite{princen0,princen00}, droplets are monodisperse
cylindrical objects packed in an hexagonal array ($n = 6$).
In the dry foam limit, this configuration satisfies Plateau's
rules for packing in two dimensions.
For $\varphi > \varphi_c$, the osmotic pressure
is given by Eq.~\ref{ae:osmotic}, with
$n = 6$~\cite{hutzler,princen00}.
For the same range, the static
shear modulus was found to obey~\cite{princen0}
\begin{equation}
G/(\sigma/R) = 0.525 \varphi^\frac{1}{2}.
\end{equation} 
Thus, this model predicts a discontinuity of $G$ at $\varphi_c$,
its value jumping from zero to $0.525\sqrt{\varphi_c}$.
The onset of the shear modulus is thus very sharp at $\varphi_c$,
and we shall demonstrate that this elastic response is intimately
associated with the response of the droplet potential
at $\xi \rightarrow 0$.

\section{Compression by two plates}
\label{s:compression}

\begin{figure}[tbp]
  \centerline{\psfig{figure=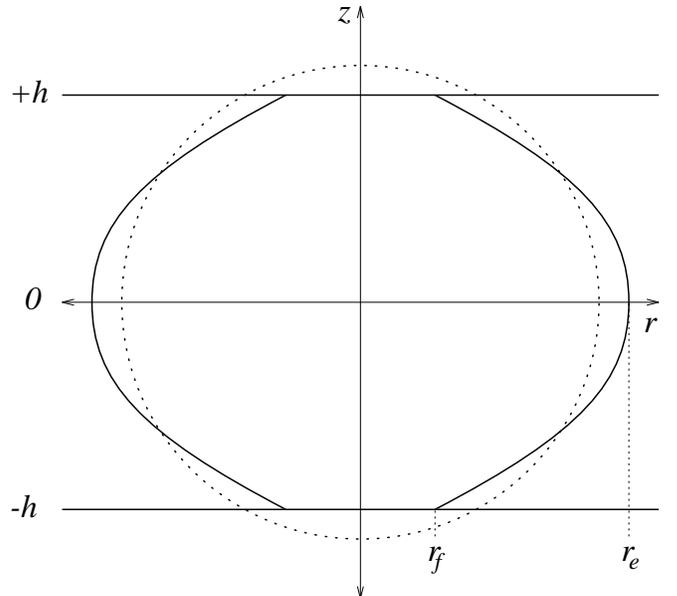,width=\figwidth}}
  \caption{A droplet of initial radius $R$
  being deformed by 2 parallel plates
  located at $z = -h$ and $z = +h$. The equator is located
  at $z=0$ and has a radius $r_e$. Both facets at
  $z = \pm h$ are circular and have a radius $r_{\!f}$. The droplet shape
  is a cubic spline drawn for visualization purposes only.
  \label{cf:schema}}
  \end{figure}

The evaluation of the excess energy stored in an arbitrary
surface deformation of a compressed droplet is a
difficult problem. A simple three-dimensional case
consists of a droplet of radius $R$
compressed between two parallel planes, each located at a distance
$h$ from the center of the droplet ({\it cf}. Fig.~\ref{cf:schema}).
We shall measure compression by the dimensionless ratio
$\xi$ defined as above.
At small compression $d\xi$,
small circular facets of radius $r_{\!f}$
appear where the droplet touches each plate.
Naively, the resulting force $dF$ on each facet can be
estimated by assuming that the radius of the droplet, and hence the
Laplace pressure, remains unchanged. Therefore,
$dF \approx (2\sigma/R)dS$,
where $dS$ is the surface of the flattened facet. To lowest
order in the deformation,
\begin{equation}
dS \approx 2\pi R^2 d\xi,
\label{ce:approx}
\end{equation}
so that
$dF \approx 4\pi\sigma R d\xi$.
Thus, to lowest order, the response of a droplet
compressed between 2 parallel plates is found to be identical to
the compression of a repulsive harmonic spring of
spring constant $4\pi\sigma$. However,
this simple derivation leads to a wrong answer, as we now show.

Because of its azimuthal symmetry, we may express this problem as a
one-dimensional problem, that of finding the solid of
revolution of a constant volume with minimum surface area. The
free surface of this solid is given by a curve $r(z)$
rotated about the z-axis.
The mathematical aspects of the present problem,
such as stability and the existence of a solution
as a function of contact angles,
have recently been discussed in detail
by various authors~\cite{gilette,michael,vogel,zhou}.

Using the Euler-Lagrange formalism,
we minimize the total droplet surface, $A$, which
is given by
\begin{equation}
A = 2 \pi r_{\!f}^2 + 2 \int_{-h}^{0} 2 \pi r \sqrt{1 + r_z^2} \:dz
\label{ce:surface}
\end{equation}
where
$r_z \equiv dr/dz$.
The minimization is done under the constraint of constant
volume:
\begin{equation}
\frac{4}{3}\pi R^3 = 2\int_{-h}^{0} \pi r^2 \:dz,
\label{ce:volume}
\end{equation}
which is introduced through the use of a Lagrange multiplier $\lambda$
in a function ${\cal L}$,
\begin{equation}
{\cal L} = \frac{r_{\!f}^2}{h} + 2 r\sqrt{1 + r_z^2} - \lambda r^2
\label{ce:function}
\end{equation}
to be minimized.
Since ${\cal L} = {\cal L} (r, r_z)$ is independent of $z$,
one can use the integrated form of the Euler-Lagrange equation~\cite{weinstock}
\begin{equation}
r_z\frac{\partial {\cal L}}{\partial r_z} - {\cal L} = C,
\label{ce:euler}
\end{equation}
where $C$ is an integration constant.
We re-express the unknown constants $\lambda$ and $C$ using the
following boundary conditions:
\begin{mathletters}
\begin{eqnarray}
\left. r_z \right|_{r_{\!f}} &=& \infty,\\
\left. r_z \right|_{r_e} &=& 0,
\end{eqnarray}
\end{mathletters}
where $r_e$ is the maximum value of $r$ located at the equator
({\it cf.} Fig.~\ref{cf:schema}). The first condition
sets the contact angle at the facet,
which should be zero if no long-range attractive
forces are present~\cite{princen0,princen1}.
The second comes from symmetry and macroscopic smoothness at the equator.

  \begin{figure}[tbp]
  \centerline{\psfig{figure=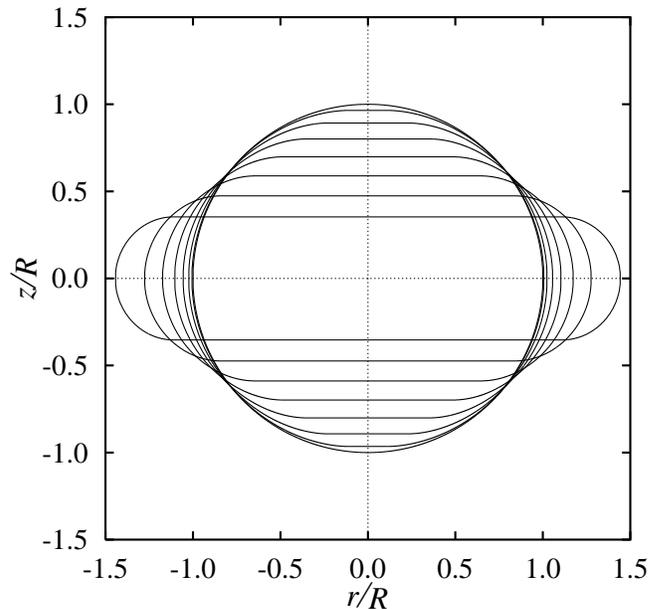,width=\figwidth}}
  \caption{Shape of a droplet compressed between two parallel plates
  for different plate displacements:
  $\xi$ = 0.000, 0.035, 0.107, 0.199, 0.302, 0.411, 0.526, 0.647.
  \label{cf:shapes}}
  \end{figure}

It will be convenient
to use the following dimensionless variables
\begin{mathletters}
\begin{eqnarray}
\rho &=& r/r_e, \\
\rho_{\!f} &=& r_{\!f}/r_e, \\
\zeta &=& z/r_e,
\end{eqnarray}
\end{mathletters}
and, including these changes, Eq.~\ref{ce:euler} reads
\begin{equation}
\sqrt{1 + \rho^2_\zeta} =
\frac{\rho ( 1 - \rho^2_{\!f})}{\rho^2 - \rho^2_{\!f}}.
\label{ce:core}
\end{equation}
Solving for $z$ and integrating, we get
\begin{equation}
z = -h + r_e I(\rho),
\label{ce:zofh}
\end{equation}
where $\rho$ goes from $\rho_{\!f}$ to 1,
and $I(\rho)$ is an integral that can
be solved at least numerically, and which is given by
\begin{equation}
I(\rho) = \int_{\rho_{\!f}}^{\rho}
\frac{x^2 - \rho_{\!f}^2}{\sqrt{(1-x^2)(x^2 - \rho_{\!f}^4)}} \:dx.
\label{ce:I}
\end{equation}
Setting $z = 0$ in Eq.~\ref{ce:zofh} we obtain a relation between
$h/r_e$ and $\rho_{\!f}$:
\begin{equation}
h = r_e I_1
\label{ce:sidekick}
\end{equation}
where $I_1 \equiv I(1)$. This allows us to re-express Eq.~\ref{ce:zofh}
as
\begin{equation}
z = -h \left[1 - I(\rho)/I_1\right].
\label{ce:z}
\end{equation}

We note here that both $I(\rho)$ and $I_1$
are functions of $\rho_{\!f}$, which in turn
is related to $h$ through the
volume constraint, Eq.~\ref{ce:volume}.
Using Eqs.~\ref{ce:volume} and~\ref{ce:sidekick}, $h$ can be related
to $\rho_{\!f}$ through the following non-trivial relation
\begin{equation}
h = R I_1 \left(\frac{2}{3J_1}\right)^{1/3},
\label{ce:h}
\end{equation}
where $J_1$ is another integral defined by
\begin{equation}
J_1 = \int_{\rho_{\!f}}^{1}
\frac{x^2(x^2 - \rho_{\!f}^2)}{\sqrt{(1-x^2)(x^2 - \rho_{\!f}^4)}} \:dx.
\label{ce:J}
\end{equation}

  \begin{figure}[tbp]
  \centerline{\psfig{figure=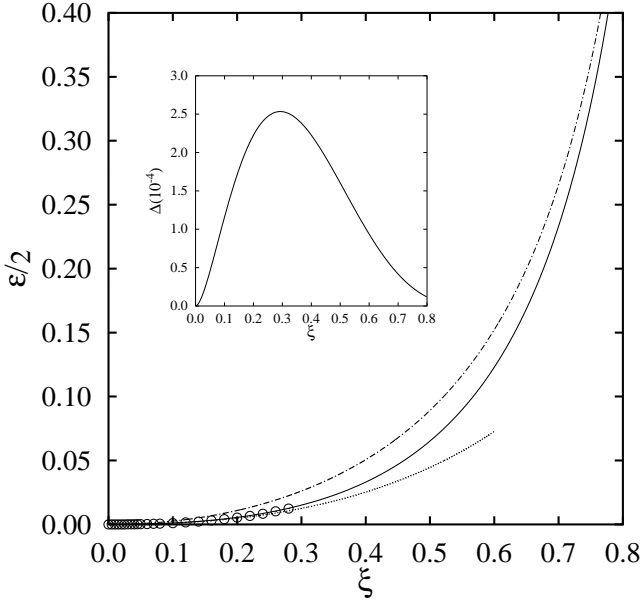,width=\figwidth}}
  \caption{Relative excess energy per facet
  for a single droplet as a function
  of the dimensionless displacement $\xi$.
  Curves are, from top to bottom:
  the truncated sphere approximation (dot-dashed),
  the exact solution (solid), and the potential
  predicted by Eq.~\protect\ref{ce:selfmorse} (dotted).
  Data points ($\circ$) are results obtained
  from SE\@.
  The absolute difference $\Delta$ in $\varepsilon$
  between the semi-circular surface
  approximation and the exact solution is shown as an insert.
  \label{cf:excess}}
  \end{figure}

At this point the problem is fully determined. Given the size of
the facets $r_{\!f}$,
the plate separation $2h$ is determined by Eq.~\ref{ce:h},
and the shape of the droplet
by Eq.~\ref{ce:z}.
The total droplet surface is obtained by combining
Eqs~\ref{ce:surface}, \ref{ce:core}, and~\ref{ce:sidekick},
to get
\begin{equation}
A = \frac{2\pi h^2}{I_1^2} \left[ \rho_{\!f}^2 + 2(1-\rho_{\!f}^2) K_1\right],
\label{ce:energy}
\end{equation}
where $K_1$ is yet another integral defined by
\begin{equation}
K_1 =
\int_{\rho_{\!f}}^{1} \frac{x^2}{\sqrt{(1-x^2)(x^2 - \rho_{\!f}^4)}} \:dx.
\end{equation}

The integrals $I_1, J_1$, and $K_1$ are
elliptic integrals which we can solve numerically,
after removing the divergence at $x = 1$ by direct
analytical integration.
The solution of Eq.~\ref{ce:z}, for various values of $h$,
leads to the different droplet shapes
shown in Fig.~\ref{cf:shapes}.

It is convenient to define the
(dimensionless) relative droplet excess surface energy by
\begin{equation}
\varepsilon \equiv A/(4\pi R^2) - 1.
\label{ce:epsdef}
\end{equation}
Figure~\ref{cf:excess} shows the relative excess energy as derived from
Eq.~\ref{ce:energy}. It is interesting to compare our solution to
different estimates, which model the deformed droplets in different
ways. For example, we consider the
relative excess energy of a body of equal volume consisting
of a cylinder of radius $r_{\!f}'$ surrounded by a semi-circle
of revolution of radius $h$. This solution has a zero contact angle
for all compressions and becomes exact in the two
displacement limits $\xi \rightarrow 0$,
and $\xi \rightarrow 1$. Given the scale of Fig.~\ref{cf:excess},
it is not possible to distinguish the results of this estimate
from the exact solution, and we thus show the difference as an insert.
A second estimate which has been used in the case of compression
by multiple planes~\cite{buzza,princen1} consists in considering
a truncated sphere having the same volume.
Results from this estimate are also shown in Fig.~\ref{cf:excess}.
The force required to compress the plates
is a better accuracy indicator, and forces derived from each solution
will be compared below.

The integrals $I_1, J_1$ and $K_1$
can be solved analytically when
$\rho_{\!f}$ = 0, i.e., for the undeformed sphere.
Let us consider small compressions, and expand the
integrals about $\rho_{\!f}$ = 0.
For $K_1$, for instance, we obtain
\begin{equation}
K_1 \approx \int_{\rho_{\!f}}^1
\left(\frac{x^2}{x\sqrt{1-x^2}} + \frac{\rho_{\!f}^4}{2}
\frac{x^2}{x^3\sqrt{1 - x^2}}\right) \:dx,
\end{equation}
which becomes, after performing the integrals,
\begin{equation}
K_1 \approx 1 - \frac{\rho_{\!f}^2}{2} + \frac{\rho_{\!f}^4}{2}
\left[\ln(2) - \frac{1}{4} - \ln(\rho_{\!f})\right].
\end{equation}
In this limit one finds,
after expanding Eq.\ref{ce:energy} in a similar way,
\begin{equation}
\varepsilon \approx -4 X^2 \left( \ln(X) + \frac{1}{2} \right),
\label{ce:dasim}
\end{equation}
where
\begin{equation}
X \equiv (\rho_{\!f}/2)^2
\label{ce:xdef}
\end{equation}
has been introduced for simplicity.
Using similar
expansions in Eq.~\ref{ce:h},
one obtains
\begin{equation}
\xi \approx -2 X \ln(X).
\label{ce:dxsim}
\end{equation}
Due to this non-trivial relation,
we cannot express $\varepsilon$
in terms of $\xi$. It is possible, however, to
combine Eqs.~\ref{ce:dasim} and~\ref{ce:xdef} to obtain
an expression for the dimensionless force~$f$
\begin{equation}
f \equiv \frac{d\varepsilon}{d\xi} \approx 4X.
\label{ce:lucky}
\end{equation}

\begin{figure}[tbp]
  \centerline{\psfig{figure=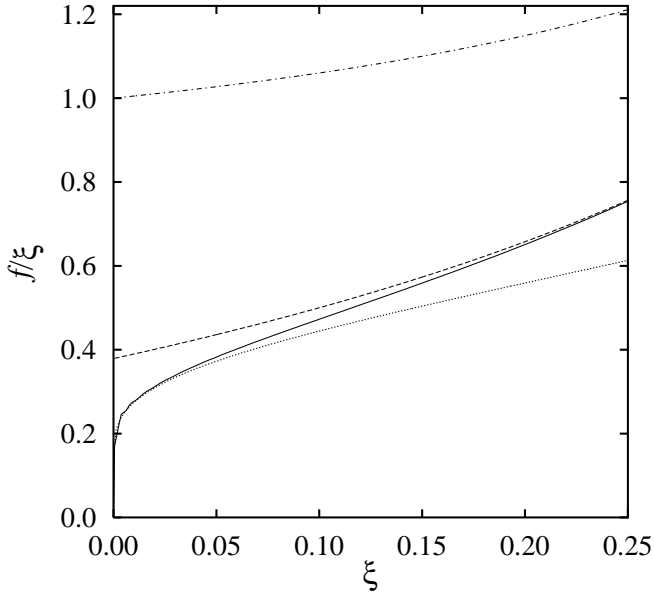,width=\figwidth}}
  \caption{A comparison of the $\xi$ dependence
  of the dimensionless force ratio $f/\xi$ for different solutions.
  Curves are, top to bottom:
  the truncated sphere approximation (dot-dashed),
  the semi-circular surface approximation (dashed),
  the exact solution (solid), and its lowest-order expansion (dotted).
  \label{cf:force}}
  \end{figure}

Combining the last equations, we may re-express the excess
energy in terms of $f$:
\begin{equation}
\varepsilon \approx \frac{1}{4}
f^2 \left[ k - \ln(f)\right],
\label{ce:final}
\end{equation}
where the constant $k = 2\ln(2) - \frac{1}{2}$.
Morse and Witten~\cite{morse} obtained a similar result by
considering, to lowest-order, a point perturbation of a sphere.
In order to test the range of validity of our expansion,
we solve Eq.~\ref{ce:final} numerically
to obtain $\varepsilon(\xi)$. This is done by transforming
Eq.~\ref{ce:final} into a self-consistent integral equation
\begin{equation}
f(\xi) = \left(\frac{4 \int_0^\xi f(x)dx}{k - \ln(f(\xi))}\right)^{1/2},
\label{ce:selfmorse}
\end{equation}
from which $\varepsilon(\xi) = \int_0^\xi f(x) dx$
is obtained after a few tens of iterations only.
The results are shown by the dotted lines on Figures~\ref{cf:excess}
and~\ref{cf:force}.

Thus, we showed that the response of a droplet
to compression is not harmonic, even to lowest order.
It is not possible to obtain an expression for the
potential as a function of $\xi$ in closed form, given the
presence of elliptic integrals. It is possible, however, to test
phenomenological expressions from our exact results.
For this purpose, the divergence of the excess energy at $h \rightarrow 0$
is not of particular interest given the perspective of
a droplet packing.
We shall therefore concentrate on the droplet
response to small compression.

It is instructive to compare the forces predicted
by the different estimates to the exact results.
Figure~\ref{cf:force} shows the ratio $f/\xi$
for the exact solution, its lowest-order expansion (Eq.~\ref{ce:final}),
the semi-circular surface model,
and the truncated sphere model. A harmonic potential
would be represented by a constant value in such a plot.
For $\xi \lesssim 0.001$, the exact solution sharply drops to zero as
does its lowest order expansion. On the other hand, as $\xi$ goes to zero,
the semi-circular surface model
goes to $(2 - 16/\pi^2)$,
showing that the potential
becomes harmonic as $\xi \rightarrow 0$. Finally, the truncated sphere
model is found to be closer to a harmonic potential, having a
smaller relative variation in $f/\xi$.
There are two essential characteristics of the exact force.
The first one is the anharmonic behavior of the force at
$\xi \rightarrow 0$. This feature is responsible for the logarithmic
onset of $G$ at $\varphi_c$ predicted for ordered emulsions~\cite{buzza}.
The second essential characteristic
is that $f/\xi$ increases as the displacement
increases. These important features, we believe, are responsible for
the characteristic positive curvature of $G$
observed for disordered droplet packings.

An approximate functional form for the droplet potential
would be useful
in providing estimates of the elastic properties of compressed emulsions.
For this purpose, the increase of $f/\xi$ with the displacement
can either be approximated by a power law
$\xi^\alpha$ with $\alpha > 2$ or by the addition
of higher-order terms to a harmonic potential. The former is
preferred however, both for its simplicity and because it
vanishes as $\xi \rightarrow 0$, similar to the exact solution.

  \begin{figure}[tbp]
  \centerline{\psfig{figure=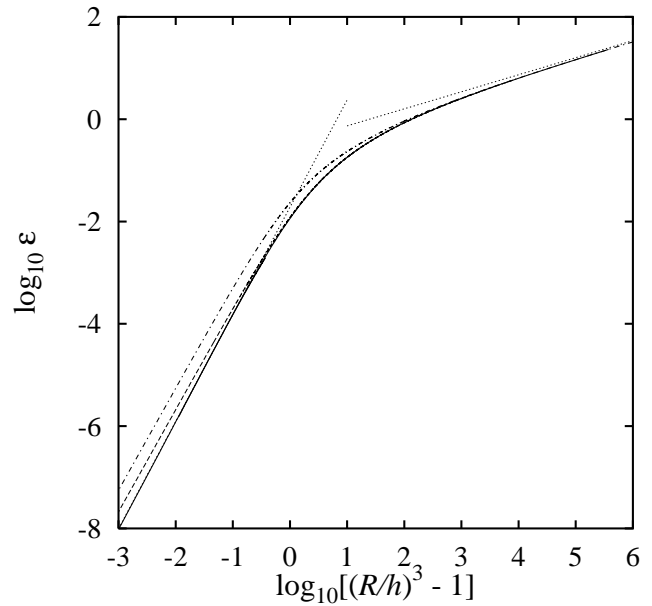,width=\figwidth}}
  \caption{Logarithmic plot of the relative
  excess energy versus the displacement function $[(R/h)^3 -1]$.
  A simple power law describes the data very well at low compression,
  as shown by the dotted line.
  The low-compression fits a slope of 2.1 while the high-compression
  data joins the asymptotic limit of slope $\frac{1}{3}$.
  Curves are, top to bottom: truncated sphere approximation (dot-dashed),
  semi-circular surface approximation (dashed), and exact (solid).
  \label{cf:logplot}}
  \end{figure}

In view of investigating fitting functions,
a logarithmic plot of $\varepsilon$
in terms of $[(R/h)^3 - 1]$ is presented in Fig.~\ref{cf:logplot}.
The numerical solution of Eq.~\ref{ce:final} can
be performed with a precision larger than the
one optimally possible when solving
directly for $\varepsilon$ (Eq.~\ref{ce:energy}).
Therefore, the low-compression values of $\varepsilon$ shown
on Fig.~\ref{cf:logplot} were obtained using the lowest-order expansion
(Eq.~\ref{ce:selfmorse}).
There are two distinct behaviors separated by a crossover section.
By approximating the droplet by
a compressed cylinder, it can easily be shown that
$\varepsilon \sim 1/h$ as $h \rightarrow 0$.
This asymptotic limit is represented
by the dotted line of slope $1/3$ at high abscissa values.
At low compression, the data are relatively
well described by a power law of the form
$\xi^\alpha$ or
$[(\frac{R}{h})^3 - 1]^\alpha$.
We found that the
latter provides a better fit over a wider range of $h$.
It also has the advantage of simplifying to the form
$(\varphi - \varphi_c)^\alpha$ for space-filling polyhedra.
Moreover, the two forms are equivalent for
small $\xi$ since $[(\frac{R}{h})^3 - 1] \approx 3\xi $.
As we shall see in the next section,
a fitting form will prove useful when characterizing
the behavior of the droplet for different
configurations of compressing planes.

As can be seen by comparing the curves of
Fig.~\ref{cf:logplot} at low compressions, a
power law describes the data very well,
down to machine precision values at $\varepsilon \approx 10^{-8}$.
The fitting curve (dotted) has an exponent $\alpha = 2.1$ and
is obtained from fitting the data over a range
covering more than an order of magnitude.
The exact solution is contrasted with the other estimates
which show a harmonic behavior $(\alpha = 2)$ at small $\xi$.
At larger $\xi$, the power law overestimates
the excess energy but
this occurs at $\xi \approx 0.2$.
Nevertheless, the range of $\xi$ over which a power law
describes the data very well is more than two decades.

  \begin{figure}[tbp]
  \centerline{\psfig{figure=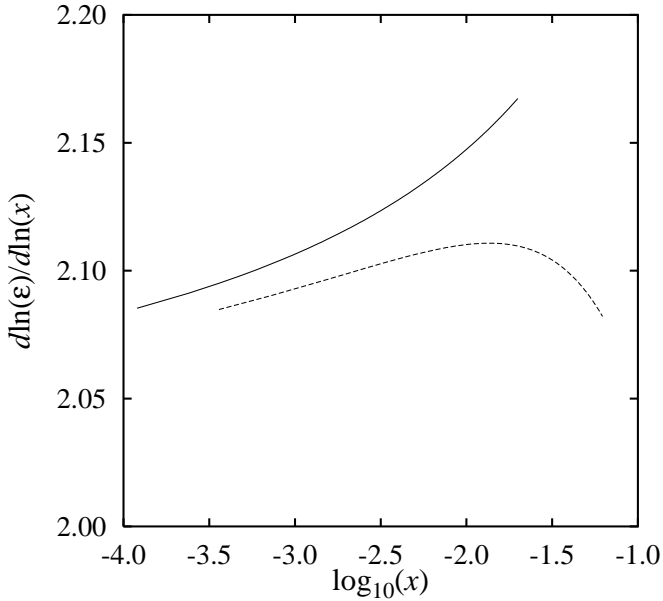,width=\figwidth}}
  \caption{The effective exponent characterizing the increase
  of the relative excess surface energy $d\ln\varepsilon/d\ln x$, where
  $x$ is either the displacement $\xi$ (solid), or a function of it
  $[1/(1-\xi)^3 - 1]$ (dashed). In both cases,
  the maximum value of $\xi$ is 2\%.
  \label{cf:alpha}}
  \end{figure}

To investigate the validity
of a power-law functional form for $\varepsilon$ at very small compressions,
we simplify Eq.~\ref{ce:final} by taking one derivative
with respect to $\xi$ and integrating $f(\xi)$ to obtain
\begin{equation}
\xi = - \frac{1}{2} f \ln(f/4).
\end{equation}
This allows us to define an effective exponent for very small compression
\begin{equation}
\frac{d\ln\varepsilon}{d\ln\xi} = \frac{\xi f}{\varepsilon}
 = \frac{2\ln(f/4)}{\ln(f/4) + 1/2},
\label{ce:alpha}
\end{equation}
characterizing the increase of the excess energy.
The exponent is obtained in terms of $f$ which can be
numerically converted to $\xi$ using Eq.~\ref{ce:selfmorse}.
Figure~\ref{cf:alpha} shows the $\xi$ dependence
of the effective exponent for the two
different fitting forms we propose here.
For both cases, points on the curve are bounded by $\xi < 2\%$,
since these results are derived from the lowest-order expansion
of the energy.
In contrast to what is suggested by Fig.~\ref{cf:logplot},
the exponent is not constant for very small compression,
but slowly decreases, going to 2 only at $\xi \rightarrow 0$.
This notwithstanding, a power law with a fixed exponent
provides a good approximation for the droplet potential
over the range of values of $\xi$ typically encountered in
compressed emulsions.

\section{Numerical results}
\label{s:evolver}

Except for special cases, as shown in the previous section, the
excess surface energy of a deformed droplet can only be determined
numerically. Given the Surface Evolver software written
by Brakke~\cite{brakke}, it is relatively easy
to calculate the
shape of a single droplet with minimum surface
area (energy) under the constraint of fixed
droplet volume and as a function of confinement.
We have obtained results for various confining polyhedral cells,
some of which are Wigner-Seitz cells associated with standard
packing structures. The surprising result of this study is that
for a considerable range of compressions, the energy is well described
by a power law, {\it where the exponent depends on the number of
faces of the confining cell.}

The starting state is a spherical droplet which is composed of either
3074 or 12290 vertices. Most of the results presented
here are for the latter case. The initial droplet is confined inside a
polyhedral cell having between four and twenty faces. In
particular, we used a rhombic dodecahedron
(face-centered cubic), a truncated octahedron (body-centered cubic)
and a simple cube (simple cubic),
all of which are space-filling polyhedra. In addition, we compressed
the droplet within
a tetrahedron, an octahedron, a pentagon dodecahedron, and an icosahedron.
Simpler structures were used as well, namely compression between
a pair of parallel plates to compare to the results
obtained in Sec.~\ref{s:compression}, and between
two perpendicular pairs of parallel plates.
Results for two parallel plates are shown as
the open circles in Fig.~\ref{cf:excess}. As expected,
they are in agreement with the analytic calculations down
to very low compression where the change in surface
area is comparable to the numerical uncertainty in SE.
In all cases, the compressions leave
the center of mass unchanged so that a
distance $h$ from the center can be defined.

  \begin{figure}[tbp]
  \centerline{\psfig{figure=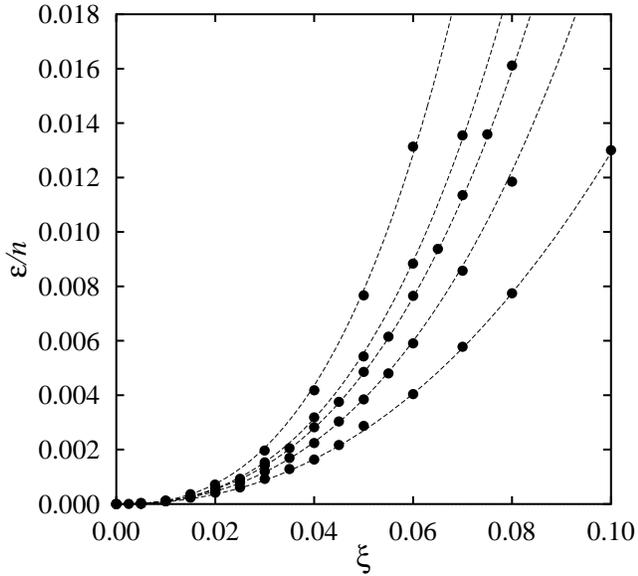,width=\figwidth}}
  \caption{Relative excess surface energy per
  facet, Eq.~\protect\ref{ce:epsdef}, for a single droplet as a function
  of the plate displacement, $\xi = (R - h)/R$.
  Curves are, from top to bottom,
  $n$ = 12 (f.c.c.), 8 (low compression b.c.c.), 6 (s.c.),
  4 (tetrahedron), 2 (2 plates).
  All curves are best fit to power law
  Eq.~\protect\ref{de:powerfit}.
  Note that for clarity, not all cases of
  Table~\protect\ref{dt:fits} are shown here.
  \label{df:excess}}
  \end{figure}

Figure~\ref{df:excess} shows the calculated dependence
of the relative excess energy (Eq.~\ref{ce:epsdef})
per facet, $\varepsilon/n$,
on $\xi$ (Eq.~\ref{ce:xidef}) for different types of confinement
cells. $n$ represents the number of compressing planes which is
equivalent to the coordination number for space-filling cells.
Using a fitting function of the form
\begin{equation}
\varepsilon/n = C \left[\left(\frac{R}{h}\right)^3 - 1\right]^\alpha,
\label{de:powerfit}
\end{equation}
we find that
both the coefficient, $C$, and the
exponent, $\alpha$, depend on $n$, as indicated in Table~\ref{dt:fits}.
As mentioned before, Eq.~\ref{de:powerfit}
is equivalent to $C'(\varphi - \varphi_c)^\alpha$ when
the droplet is confined by space-filling
polyhedra. As in the case of two parallel plates, this form fits
the data better over a wider range than $\xi^\alpha$.
The data selected for the fit lie in the $\xi$ interval 2--6\%.
This range was chosen to avoid errors associated with low values
of $\xi$. On the other hand, the upper value fixes the range we
are interested in.
Indeed, since $\xi = 1 - (\varphi_c/\varphi)^{\frac{1}{3}}$,
a face-centered-cubic structure (f.c.c., $\varphi_c = \pi\sqrt{2}/6$)
for instance,
has $\xi \approx 9.5\%$ in the dry foam limit $\varphi = 1$.

Our results explicitly show that the response
of a droplet to compression is a non-local phenomenon: {\it the response
depends on the number of planes used to compress the droplet}. In
addition, one might expect it to depend on the relative positions
of these planes, but it will be shown below
that this effect is not very important. It is worth noting that since
$\alpha(n) > 2$ for all cases, this
response is weaker than a harmonic interaction: the response
is equivalent to having the spring
constant go to zero as the distortion vanishes.
We also note that a power law with an exponent which increases as the
number of compressing planes increases is unphysical for small
compressions as the energy curves for different $n$ would cross.
Thus, although the fits
shown in Fig.~\ref{df:excess} appear convincing,
the functional form should be viewed
uniquely as a convenient way to mimic the response of a droplet.

\begin{table}[tbp]
\caption{Values of the parameters $C$ and $\alpha$ of the power-law fit to
the relative excess energy (Eq.~\protect\ref{de:powerfit}).
The various configurations are ordered according to
their coordination number $n$. Errors are in the last digit.}
\begin{tabular}{lccc}
configuration & $n$ & $C (10^{-3})$ & $\alpha$ \\ \tableline
pair of $\parallel$ plates & 2 & 7 & 2.1 \\
two $\perp$ pairs of $\parallel$ plates & 4 & 12 & 2.1 \\
tetrahedron & 4 & 14 & 2.1 \\
cube (s.c.) & 6 & 21 & 2.2 \\
octahedron (low-$\varphi$ b.c.c.) & 8 & 27 & 2.3 \\
rhombic dodecahedron (f.c.c.) & 12 & 55 & 2.5 \\
pentagonal dodecahedron & 12 & 54 & 2.5 \\
icosahedron & 20 & 69 & 2.6
\end{tabular}
\label{dt:fits}
\end{table}

In view of investigating the dependence of the relative positions
of the compressing planes on the droplet, we obtained data for different
configurations having the same $n$.
By comparing, say, the results of the f.c.c.\ structure to
those of the pentagon dodecahedron, we see that the influence
of the configuration on the fitting parameters is marginal
compared to that of $n$.
It therefore seems reasonable to conclude that, provided that
the different facets of the droplet are not too close, the potential
is only a function of the coordination number $n$.
This information is valuable for modeling
the excess energy of a droplet in a disordered
emulsion~\cite{lacasse}.
A second point of interest is the saturation of the exponent $\alpha$
above $n\approx 12$. Although the number of neighbors nearly doubles
from a rhombic dodecahedron to an icosahedron, the exponent
barely changes.

In a similar way, by comparing the excess energy obtained by
compressing a droplet in an octahedron to
that obtained for the truncated octahedron,
one can determine the importance of the second neighbors
in the base-centered-cubic (b.c.c.) lattice.
We find that second neighbors do not
play a role for $\varphi \lesssim 0.90$ ($\xi \approx 6.3\%$).
For $\varphi=0.93$, the excess energy differences
between the two is about 5\%. Thus over most of the range
of interest, only the eight nearest neighbors are
relevant.

\section{Optimal packings for foam}
\label{s:foam}

Our numerical results also allow us to compare the energy of the
various packings as a function of volume fraction.
For a wide range of volume fractions,
the lowest energy state of monodisperse emulsions
is believed to be a packing of droplets in an
f.c.c.\ structure. We neglect temperature effects in
the following argument.
For $\varphi > \varphi_m \approx 0.545$, i.e.\ the melting
concentration of monodisperse hard spheres~\cite{hoover},
up to $\varphi < \varphi_c = \pi\protect\sqrt{2}/6 \simeq 0.7405$,
i.e.\ close packing concentration,
the droplets remain spherical in an f.c.c.\ structure
as they can pack without touching.
For $\varphi > \varphi_c$, the lowest energy state
remains f.c.c.\ until it changes to a new structure
at $\varphi^*<1$.
That such a crossover must exist is clear from previous
work on dry foams.

In the limit $\varphi \approx 1$, an emulsion becomes a biliquid foam which
is structurally analogous to a dry foam.
The lowest-energy state of the latter
has long been the object of study. For nearly one hundred years,
it was thought that the Kelvin~\cite{kelvin} structure
based on a b.c.c.\ packing of identical
orthic tetrakaidecadedra gave an optimal space filling
of cells of the same volume.
The mechanical equilibrium requirements of the foam structure
are contained in Plateau's rules~\cite{plateau}.
To satisfy these rules, Kelvin showed that the
faces of each cell are slightly distorted~\cite{kelvin}:
the six quadrilateral surfaces remain planar but curve their edges
while the eight hexagonal surfaces, in addition to sharing
the same curved edges, become non-planar surfaces of
zero mean curvature. The reduction in surface area from the
undistorted Wigner-Seitz cell is approximately $0.16\%$~\cite{princenl,reinelt}.
By using SE, Weaire and Phelan~\cite{weaire} recently showed the existence
of a packing having an energy lower than the Kelvin structure.
This space-filling packing consists of
six 14-sided polyhedra and two 12-sided polyhedra,
all of equal volume, packed in a simple cubic cell.
This new structure, which is
based on the cubic clathrate structure, has a surface energy
approximately $0.3\%$ less than that of Kelvin.
For comparison,
the f.c.c.\ structure has an energy $0.7\%$ higher than the Kelvin
structure~\cite{weaire}.

  \begin{figure}[tbp]
  \centerline{\psfig{figure=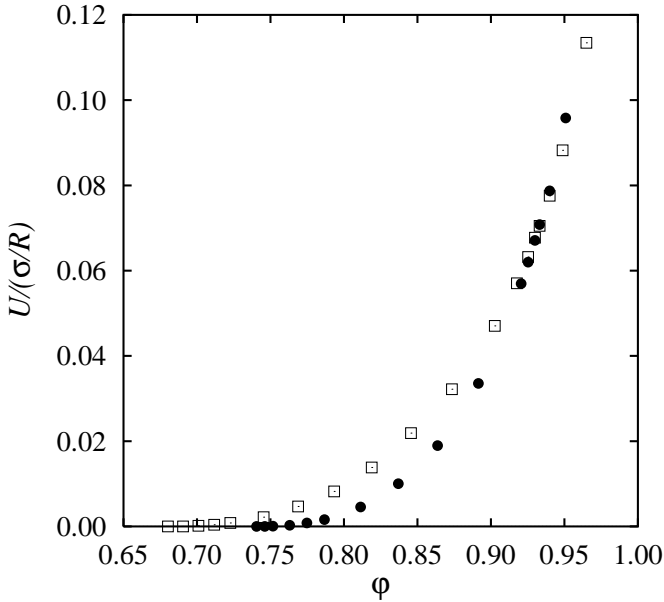,width=\figwidth}}
  \caption{The scaled excess energy density versus the
  volume fraction as extracted from a droplet
  in a b.c.c.\ $(\Box)$ and f.c.c.\ $(\bullet)$ lattice.
  We find a crossover from f.c.c.\
  to b.c.c.\ at $\varphi^* = 0.932(1)$.
  $\varphi_c = \pi\protect\sqrt{3}/8 \simeq 0.6802$ for b.c.c.\ and
  $\pi\protect\sqrt{2}/6 \simeq 0.7405$ for f.c.c.
  \label{df:prediction}}
  \end{figure}

Since we have studied only a single droplet in a Wigner-Seitz
cell, we cannot address the stability of the Kelvin structure
versus that of Weaire-Phelan for $\varphi <1$. However, we can
study the crossover from the f.c.c.\ packing to the b.c.c.\
Kelvin-like packing as $\varphi$ increases. In Fig.~\ref{df:prediction},
we plot the (scaled) energy density, $U$, defined by
\begin{equation}
U = E/V = 3 \varphi\varepsilon (\sigma/R)
\label{de:dedef}
\end{equation}
as a function of $\varphi$ for a droplet in
b.c.c.\ and f.c.c.\ structures.
Here, $E$ is the excess surface energy $E = \sigma (A - 4\pi R^2)$.
As expected, the energy density of the
f.c.c.\ structure is the lowest over a wide range
of $\varphi$. But at $\varphi^* = 0.932(1)$, we
observe a crossover to the b.c.c.\ structure.
This result is in agreement with a
similar prediction by Kraynik~\cite{kraynik2}.
Note that this crossover value is obtained
strictly from surface energy
considerations and may be superseded by other stability criteria.
Moreover, our calculation neglects some degrees of freedom
since it strictly involves compressing a
droplet inside a polyhedral cell. This is more restrictive
than compressing many droplets caged by their neighbors, and thus
we are in fact obtaining an upper bound on the energy.
However, since the energy of a drop in the shape of a
truncated octahedron is only
$0.16\%$ higher than the orthic tetrakaidecadedra of the
Kelvin structure in the dry foam
limit, the corrections to the crossover
value resulting from the surface curvature of the films
between the droplets should be small. Note that the
crossover point is just above the volume fraction
where second neighbors begin to play a role
in the energy of a compressed droplet in the b.c.c.\ phase.
Thus, while the six second neighbors contribute to increasing the
deformation of a droplet and thereby the surface area, the overall
energy of the b.c.c.\ structure is still lower than that of the
f.c.c.\ structure for $\varphi>0.932$. As we shall see in the next section,
second neighbors in the b.c.c.\ structure are required
for stability under shear deformation.

The little difference in energy shown in Fig.~\ref{df:prediction}
is not specific to b.c.c.\ and f.c.c.\
pairs. In fact, it is interesting to compare the energy per
droplet obtained for the regular pentagon dodecahedron (r.p.d.)
to those obtained for the b.c.c.\ and f.c.c.\ lattices.
Since the r.p.d.\ is not a space-filling object,
$\varphi$ represents the ratio of droplet to cell volume for this case.
If one compares the energy per droplet as a function of the
volume ratio, one finds that the results for the r.p.d.\
are almost indistinguishable from that of a
truncated octahedron, i.e.\ the b.c.c.\ unit cell. In particular, results
for the r.p.d. are lower than that for f.c.c.\ at high $\varphi$.
This suggests that it may be favorable for a
compressed monodisperse emulsion to generate
short-range icosahedral order such as it
exists in metallic glasses. This implies that as an
emulsion is compressed, it is likely to
undergo a glass transition reminiscent
of that occurring in metallic glasses. Indeed, it is found to be
very difficult to obtain an ordered monodisperse emulsion experimentally,
indicative of the deep energy minimum of an amorphous state.

\section{Response to shear}
\label{s:shear}

We now turn to the response of a compressed droplet to an impressed
shear strain. Since we are considering a single droplet,
the scope of our method is limited to ordered lattices.
All the space-filling confinement cells investigated here are related to the
cubic point group and consequently, the associated elastic constant tensor
is composed of three independent quantities. It is not possible
for us to determine these constants individually.

A useful method for applying a shear strain in
presence of periodic boundary conditions is
to use an isochoric uniaxial compressional-extensional
strain. The effective shear modulus can then be extracted
from the resulting change in surface energy.
The isochoric uniaxial strain consists in an extension of the Wigner-Seitz
cell of $\lambda = 1 + \epsilon$ in the, say, $z$ direction, associated
with a compression of $\lambda^{-\frac{1}{2}}$ in the perpendicular plane.
In principle, different orientations of the Wigner-Seitz cell
with respect to the reference $z$ axis would lead to effective
measures of the modulus resulting from different
combinations of the elastic constants.
However, considerations of the alignment of the facets through
periodic boundary conditions put
restrictions on the possible orientations of the strains.
Or, said differently, some orientations of the uniaxial
strain involve forces that are not normal
to the faces of the Wigner-Seitz cell, and it results
that some facets do not lie on the cell boundary~\cite{buzza}.

  \begin{figure}[tbp]
  \centerline{\psfig{figure=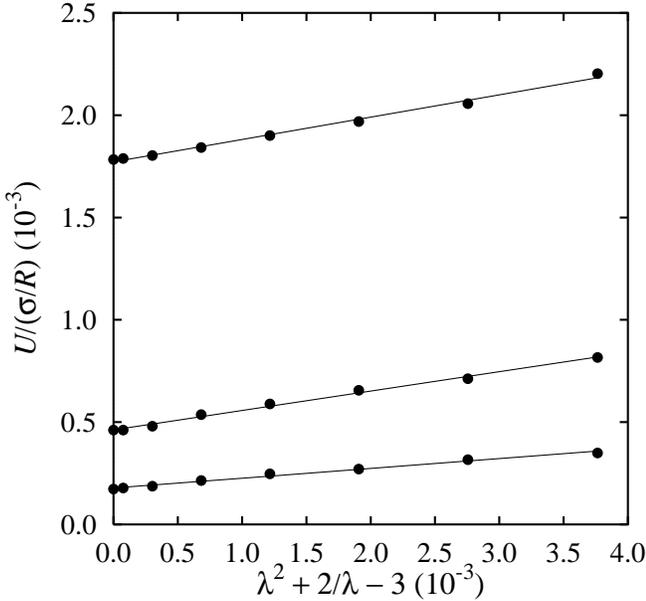,width=\figwidth}}
  \caption{Typical calculated excess energy density of a droplet
  compressed in an f.c.c.\ lattice as a function
  of shear strain. The strain is measured by
  $\lambda^2 + 2/\lambda - 3 \approx 3\epsilon^2$ so
  that the slopes are half the values of shear moduli~\protect\cite{treloar}.
  Curves are for, top to bottom, $\varphi$ = 0.787, 0.763, 0.746.
  \label{ef:strain}}
  \end{figure}

Here, the chosen orientation is along any of the three
natural directions
of a unit cell. For this particular choice, the strain tensor
is written~\cite{bisplinghoff}
\begin{equation}
\gamma_{mn} = \frac{1}{2} \delta_{mn} (\lambda_m^2 - 1),
\end{equation}
where $\delta_{mn}$ is the Kronecker delta,
$\lambda_1 = \lambda_2 = \lambda^{-\frac{1}{2}}$, and
$\lambda_3 = \lambda = 1 + \epsilon$. $\epsilon$
may be thought of as the magnitude of the small linear strain.
The energy density $U$ of the structure is given by
\begin{equation}
U = \frac{1}{2} \gamma_{mn}E_{mnpr}\gamma_{pr},
\label{ee:hook}
\end{equation}
where we use the summation convention.
$E$ is a fourth-order tensor which can be displayed as
a $6 \times 6$ coefficient matrix using Voigt notation~\cite{buzza,landau7}.
When a small isochoric uniaxial strain is applied,
the energy density is found to vary as
\begin{equation}
U = \frac{3}{2} G \epsilon^2,
\end{equation}
where the effective shear modulus $G$ is the combination
of the elastic constants $(C_{11} - C_{12})/2$.
Our definition of $G$ reduces consistently to the
corresponding Lam\'e constant
when considering an isotropic material.

Figure~\ref{ef:strain} shows a typical result for the variation
of surface area as a function of shear strain for an f.c.c.\ lattice
at different volume fractions.
The strain is measured as $\lambda^2 + 2/\lambda - 3$
which is equivalent to $3\epsilon^2$ for
small strains~\cite{treloar}. The slope of each curve
is $G/2$.

\begin{figure}[tbp]
  \centerline{\psfig{figure=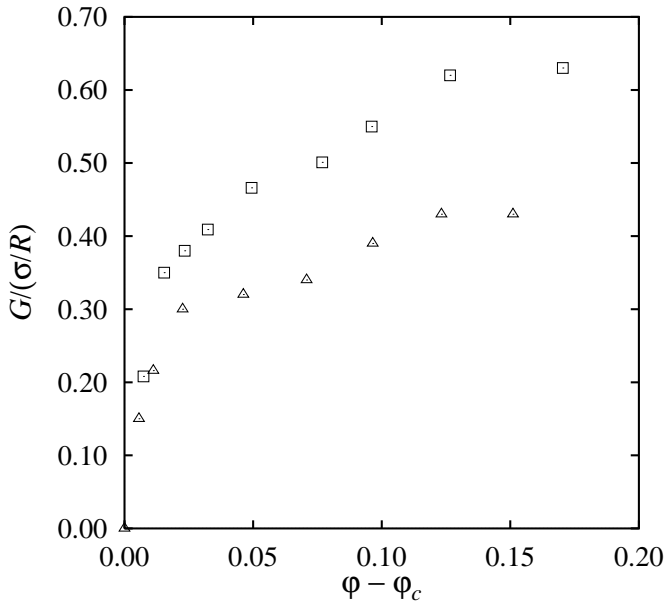,width=\figwidth}}
  \caption{Shear modulus for a droplet in s.c.\ ($\Box$) and f.c.c.\
  ($\triangle$) Wigner-Seitz confinement cells as a function
  of compression. Errors are of the order of 10\%.
  $\varphi_c = \pi/6 \simeq 0.5236$ for s.c.\ and
  $\pi\protect\sqrt{2}/6 \simeq 0.7405$ for f.c.c.
  \label{ef:modulus}}
  \end{figure}

In Fig.~\ref{ef:modulus}, the shear modulus is shown for the
s.c.\ and f.c.c.\ lattices as a function of volume fraction.
These results compare well
with the ones obtained by Buzza and Cates~\cite{buzza}
using Morse and Witten's potential~\cite{morse}.
Similar to what they found for a s.c. lattice, the shear modulus
of the f.c.c.\ structure shows a smooth rise (rather than a jump)
at $\varphi_c$. As we already mentioned, this behavior
is different from that observed experimentally for disordered
monodisperse emulsions.

The isochoric uniaxial strain is peculiar in that with
$\lambda = 2^\frac{1}{3}$, it changes the metric tensor in such
way to transform a b.c.c.\ lattice into an f.c.c.\ lattice with
the same density~\cite{rahman}. The net energy required for this
transformation will depend on the volume fraction as we
have seen in the previous section when comparing the energy
of both lattices.
For $\varphi > \varphi_c$ and up to some $\varphi' < \varphi^*$,
the b.c.c.\ structure was found to be unstable ($G<0$) to an applied
uniaxial strain. The existence of $\varphi'$ derives
from the stability of the b.c.c.\ lattice over the
f.c.c.\ lattice for $\varphi > \varphi^*$.
Second neighbors are responsible for stabilizing the structure
against shear strain and it is not surprising to find
that $\varphi' \gtrsim 0.90$, the point where second neighbors
start touching the droplet. Our best estimate for $\varphi'$,
i.e.\ the volume fraction above which a b.c.c.\ lattice has a
positive shear modulus for an isochoric uniaxial strain, is 0.903(5).

\section{Discussion and Conclusion}
\label{s:discussion}

We have shown that the droplet response to small compressions in
three dimensions is not
harmonic and that it depends on the number of neighbors.
For droplets compressed between two parallel plates, the surface
profile can be solved analytically for small deformations and integrated
numerically for arbitrary deformations. However for other geometries,
the deformation can only be determined numerically. Using the
Surface Evolver software~\cite{brakke}, we have determined the
change in surface area of droplets compressed in a variety
of confining polyhedra.
Our results strongly suggest that
in three dimensions, though not in two, the response is non-local.
The data fit very well with a non-harmonic
scaling form in which the parameters of the fit depend
almost exclusively on the coordination number.
This feature has been used
elsewhere~\cite{lacasse} to build a pairwise
inter-droplet potential in order to
investigate the elastic response of disordered monodisperse
emulsions.

The response to shear was also investigated
for ordered lattices and it was found that similar to previous
studies~\cite{buzza}, the volume-fraction dependence of the shear
modulus shows a rather sharp increase at the onset of
droplet touching, $\varphi_c$. Although the anharmonicity of the
potential certainly plays a role in the linear onset of $G$ for
disordered emulsions, it was shown here that it is not sufficient.
The experimental results currently available~\cite{mason,princenka}
are for disordered emulsions only. Ordered lattices are easier to
study theoretically but the effects of disorder seem to be too
important to permit comparison with experiment.
We have shown elsewhere~\cite{lacasse} that
both anharmonicity and disorder are required to reproduce the
experimental results.

\section*{Acknowledgments}
We thank Shlomo Alexander, Wei Cheng Cai, Thomas Halsey, and
David Weitz for useful discussions, and
{\it Le Fonds FCAR du Qu\'ebec\/} and
the U.S.-Israel Binational Science Foundation for financial support.

%%%%%%%%%%%%%%%%%%%%%%%%%%%%%%%%%%%%%%%%%%%%%%%%%%%%%%%%%%%%%%%%%%%%%%%%%%
%% References
%%%%%%%%%%%%%%%%%%%%%%%%%%%%%%%%%%%%%%%%%%%%%%%%%%%%%%%%%%%%%%%%%%%%%%%%%%

%%%%%%%%%%%%%%%%%%%%%%%%%%%%%%%%%%%%%%%%%%%%%%%%%%%%%%%%%%%%%%%%%%%%%%%%%%
%% Figures
%%%%%%%%%%%%%%%%%%%%%%%%%%%%%%%%%%%%%%%%%%%%%%%%%%%%%%%%%%%%%%%%%%%%%%%%%%

%% Cut here for preprints since figures are floating in the text.

\begin{references}
  \vspace*{\refspace}

\bibitem{princen0} H.M. Princen, J. Colloid Interface Sci.
{\bf 91}, 160 (1983).

\bibitem{kraynik} A.M. Kraynik, Ann. Rev. Fluid Mech. {\bf 20}, 325 (1988).

\bibitem{bolton} F. Bolton and D. Weaire, \prl {\bf 65}, 3449
(1990).

\bibitem{morse} D.C. Morse and T.A. Witten, Europhys. Lett. {\bf 22}, 549
(1993).

\bibitem{buzza} D.M.A. Buzza and M.E. Cates, Langmuir {\bf 10}, 4503
(1994).

\bibitem{mason} T.G. Mason, J. Bibette, and D.A. Weitz, Phys. Rev.
Lett. {\bf 75}, 2051 (1995).

\bibitem{berryman} J.G. Berryman, \pra {\bf 27}, 1053 (1983)
and references therein.

\bibitem{princenka} H.M. Princen and A.D. Kiss, J. Colloid Interface Sci.
{\bf 112}, 427 (1986).

\bibitem{brakke} K. Brakke, Exp. Math. {\bf 1}, 141 (1992).

\bibitem{plateau} J. Plateau, {\it Statique Exp\'erimentale et Th\'eorique des
Li\-quides Soumis aux Seules Forces Mol\'eculaires},
(Gauthier-Villars, Paris, 1873).

\bibitem{hutzler} S. Hutzler and D. Weaire, J. Phys.: Condens. Matter
{\bf 7}, L657 (1995).

\bibitem{durian} D. Durian, \prl {\bf 75}, 4780 (1995).

\bibitem{princen1} H.M. Princen, M.P. Aronson, and J.C. Moser,
J. Colloid Interface Sci. {\bf 75}, 246 (1980).

\bibitem{lacasse} M.-D. Lacasse, G.S. Grest, D. Levine, T.G. Mason,
and D.A. Weitz, \prl {\bf 76}, xxx (1996).

\bibitem{bideau} D. Bideau and J.P. Troadec, J. Phys. C {\bf 17},
L731 (1984) and references therein.

\bibitem{princen00} H.M. Princen, J. Colloid Interface Sci.
{\bf 71}, 55 (1979).

\bibitem{gilette} R.D. Gilette and D.C. Dyson, Chem. Engrg. J. {\bf 2}, 44
(1971).

\bibitem{michael} D.H. Michael, Ann. Rev. Fluid Mech. {\bf 13},
189 (1981).

\bibitem{vogel} T.I. Vogel, SIAM J. Appl. Math. {\bf 47}, 516, (1987);
{\it Ibid.} {\bf 49}, 1009, (1989).

\bibitem{zhou} L. Zhou, Ph.D. Dissertation, Stanford University (1995);
preprints.

\bibitem{weinstock} R. Weinstock, {\it Calculus of variations}
(Dover, New York, 1974).

\bibitem{hoover} W.G. Hoover and F.H. Ree, J. Chem. Phys. {\bf 49},
3609 (1968).

\bibitem{kelvin} W. Thomson (Lord Kelvin), Phil. Mag. {\bf 24}, 503 (1887).

\bibitem{princenl} H.M. Princen and P.J. Levinson, J. Colloid Interface
Sci. {\bf 120}, 172 (1987).

\bibitem{reinelt} D.S. Reinelt and A.M. Kraynik, J. Colloid Interface Sci.
{\bf 159}, 460 (1993).

\bibitem{weaire} D. Weaire and R. Phelan, Phil. Mag. Lett. {\bf 69}, 107 (1994).

\bibitem{kraynik2} A.M. Kraynik, private communication.

\bibitem{bisplinghoff} R.L. Bisplinghoff, J.W. Mar, and T.H.H. Pian,
{\it Statics of deformable solids} (Dover, New-York, 1990).

\bibitem{landau7} L. Landau and E. Lifshitz,
{\it Theory of Elasticity} (Pergamon, Oxford, 1959).

\bibitem{rahman} A. Rahman and G Jacucci, Nuovo Cimento D {\bf 4}, 357 (1984).

\bibitem{treloar} L.R.G. Treloar, Rep. Prog. Phys. {\bf 36}, 755 (1973).

\end{references}
  \end{document}